\documentclass{article}
\usepackage{graphics}
\begin{document}
\begin{centering}
\title {Electromagnetic Waves in a Rotating Frame of Reference}
\maketitle
\vspace{28pt}
\end{centering}
\begin{centering}  John C. Hauck and Bahram Mashhoon\\

Department of Physics and Astronomy\\
University of Missouri-Columbia\\
Columbia, Missouri 65211, USA\\
\end{centering}
\vspace{2in}
\begin{center}  {\bf Abstract}
\end{center}

We discuss the electromagnetic measurements of rotating observers and study the propagation of electromagnetic waves in a uniformly rotating frame of reference.  The phenomenon of helicity-rotation coupling is elucidated and some of the observational consequences of the coupling of the spin of a particle with the rotation of a gravitational source are briefly examined.\newline
\newpage
\section {Introduction}

\vspace{.15in}
Electrodynamics of accelerated media and various electromagnetic measurements in non-inertial frames have been discussed from various points of view by a number of authors (see [1]-[6] and the references cited therein).  We are interested in the phenomena associated with the coupling of photon spin with the rotation of the observer; therefore, this paper is devoted to a detailed discussion of helicity-rotation coupling.

Imagine an observer or a measuring device $D$ at the origin of a global inertial frame in Minkowski spacetime.  The term ``observer'' will be used throughout in an extended sense to include any relevant measuring device; hence, as employed here an observer is not necessarily sentient.  In terms of inertial coordinates $x^{\mu}=(ct,x,y,z)$, the background metric is $ds^{2}=\eta_{\mu\nu}dx^{\mu}dx^{\nu}$, where $\eta_{\mu\nu}$ is the Minkowski metric tensor with signature $+2$.  Units are chosen such that $c=1$; nevertheless, we keep $c$ in many of the formulas for the sake of clarity.  Let the observer $D$ refer its observations to spatial axes that rotate uniformly with frequency $\Omega$; that is, $D$ rotates uniformly while its center of mass is at rest at the origin of spatial coordinates.  Choosing the fixed axis of rotation to be the $z$-axis, we can write the tetrad frame of the class of rotating observers at rest on the axis of rotation as
\begin{eqnarray}
\lambda^\mu_{\;\;(0)} &=& (1,0,0,0), \label{eq:tetrad} \nonumber \\
\lambda^\mu_{\;\;(1)} &=& (0,\cos{\phi},\sin{\phi},0), \\
\lambda^\mu_{\;\;(2)} &=& (0,-\sin{\phi},\cos{\phi},0), \nonumber \\
\lambda^\mu_{\;\;(3)} &=& (0,0,0,1), \nonumber
\end{eqnarray}
where $\phi = \Omega t+\varphi$.  Here $\varphi$ is a constant angular parameter for each observer; for the reference observer $D$, we choose $\varphi_D=0$.

Consider now a geodesic coordinate system established along the worldline of the fiducial observer $D$ [7].  That is, at any instant of proper time $t$ along the worldline of $D$ at event $x^{\mu}_{D}$, we consider the spacelike hyperplane orthogonal to the worldline.  Let $x^{\mu}$ be the inertial coordinates at an event on this hyperplane and $x'^{\mu}$ be the corresponding coordinates with respect to the geodesic coordinate system.  Then $x'^{0}=ct$ and $x'^{i}=(x^{\mu}-x^{\mu}_{D})\lambda^{\;\;(i)}_{\mu}$.  This geometric construction provides the justification for the use of the standard coordinate transformation $x^{\mu}\rightarrow x'^{\mu}$, where 
\begin{eqnarray}
t' &=& t, \nonumber \\
x' &=& x\;\cos{\Omega t} + y\;\sin{\Omega t}, \\
y' &=& -x\;\sin{\Omega t} + y\;\cos{\Omega t}, \nonumber \\
z' &=& z. \nonumber 
\end{eqnarray}
We are interested in the results of measurements of the class of fundamental observers (i.e. those at rest) in this rotating coordinate system.  Each such observer is noninertial and rotates uniformly with speed $\beta c=\Omega \rho$ on a circle of radius $\rho=\sqrt{x^{2}+y^{2}}$ around the $z$-axis.  The tetrad frame $\Lambda^{\mu}_{\;\;(\alpha)}$ of a fundamental observer can be obtained from (1) by a Lorentz transformation (see Figure 1)
\begin{figure}
\includegraphics{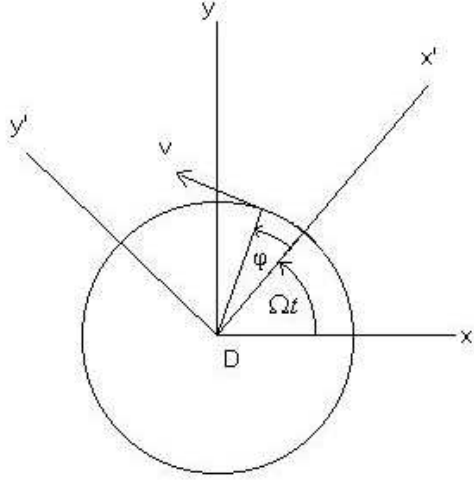}
\caption{A rotating observer with speed $v=c\beta$ at $\phi=\Omega t+\varphi$}
\end{figure}
\begin{eqnarray}
\Lambda^{\mu}_{\;\;(0)} &=& \gamma [\lambda^{\mu}_{\;\;(0)}+\beta \lambda^{\mu}_{\;\;(2)}] \nonumber \\
\Lambda^{\mu}_{\;\;(1)} &=& \lambda^{\mu}_{\;\;(1)} \nonumber \\
\Lambda^{\mu}_{\;\;(2)} &=& \gamma [\lambda^{\mu}_{\;\;(2)}+\beta \lambda^{\mu}_{\;\;(0)}] \\
\Lambda^{\mu}_{\;\;(3)} &=& \lambda^{\mu}_{\;\;(3)}. \nonumber 
\end{eqnarray}
Thus we have
\begin{eqnarray}
\Lambda^{\mu}_{\;\;(0)} &=& \gamma (1,-\beta \sin{\phi},\beta \cos{\phi},0), \nonumber \\
\Lambda^{\mu}_{\;\;(1)} &=& (0,\cos{\phi},\sin{\phi},0), \nonumber \\
\Lambda^{\mu}_{\;\;(2)} &=& \gamma (\beta,-\sin{\phi},\cos{\phi},0), \\
\Lambda^{\mu}_{\;\;(3)} &=& (0,0,0,1), \nonumber
\end{eqnarray}
where $\gamma = (1-\beta^{2})^{-1/2}$ and (4) reduces to (\ref{eq:tetrad}) for $\rho=0$.  The worldlines representing these fundamental rotating observers that are static in the rotating coordinate system fill an open cylindrical region in Minkowski spacetime around the worldline of $D$; the boundary of this region is the light cylinder given by $\rho = c/\Omega$.  The light cylinder is a timelike hypersurface; therefore, observers inside this cylinder can in principle communicate with the outside world without any difficulty.

Expressed in terms of the $(t',x',y',z')$ coordinate system, the tetrad of the fundamental observer at $x'= \rho \cos{\varphi},\;\;y'=\rho \sin{\varphi}$ and $z'=z$ is given by [8]
\begin{eqnarray}
\Lambda'^{\mu}_{\;\;(0)} &=& \gamma (1,0,0,0), \nonumber \\
\Lambda'^{\mu}_{\;\;(1)} &=& (0,\cos{\varphi},\sin{\varphi},0), \nonumber \\
\Lambda'^{\mu}_{\;\;(2)} &=& (\gamma \beta,-\gamma^{-1}\sin{\varphi},\gamma^{-1} \cos{\varphi},0), \label{eq:tetrad3} \\
\Lambda'^{\mu}_{\;\;(3)} &=& (0,0,0,1). \nonumber
\end{eqnarray}
Moreover, the spacetime metric in the rotating frame is given by \newline $ds^{2}=g'_{\mu \nu}(x')dx'^{\mu}dx'^{\nu}$, where
\begin{equation}
(g'_{\mu \nu}) = \left[ \begin{array}{cccc} -1+\Omega^{2}(x'^{2}+y'^{2}) & -\Omega y' & \Omega x' & 0 \\ -\Omega y' & 1 & 0 & 0 \\ \Omega x' & 0 & 1 & 0 \\ 0 & 0 & 0 & 1 \end{array} \right] 
\end{equation}
and the rotating coordinates are admissible within the light cylinder.  The limitations of spacetime measurements by rotating observers have been discussed in detail in [7,9,10].  In this paper, we are interested in the electromagnetic measurements of rotating observers.
Imagine an electromagnetic radiation field characterized by the Faraday tensor $F_{\mu \nu}$ in the inertial frame.  For the sake of simplicity we assume that $F_{\mu \nu}$ is complex and the actual field is Re$(F_{\mu \nu})$.  This simplifying assumption is adopted throughout this paper, since all of the field operations that we consider are linear.  The standard assumption in the theory of special relativity regarding what an accelerated observer would measure is the \emph{hypothesis of locality}, namely, the supposition that a noninertial observer is at each instant equivalent to an otherwise identical momentarily comoving inertial observer.  Along its worldline, the noninertial observer thus passes through a continuous infinity of hypothetical momentarily comoving inertial observers [10].  The decomposition of $F_{\mu \nu}$ into its Fourier components each with wave vector \textbf{k} and frequency $\omega = c |\bf{k}|$ implies that at each instant of time the noninertial observer measures
\begin{eqnarray}
\omega' &=& \gamma (\omega - {\bf{v}} \cdot {\bf{k})}, \\
{\bf{k}'} &=& {\bf{k}} + \frac {1}{v^{2}}(\gamma -1)({\bf{v}} \cdot {\bf{k}}){\bf{v}} - \frac{1}{c^{2}}\gamma\omega{\bf{v}},
\end{eqnarray}
according to the standard relativistic formulas for Doppler effect and aberration.  Since $\omega'$ and $\textbf{k}'$ in general change with time from one instant to the next and a few periods of the wave must be registered by the observer in any case before $\omega'$ and $\bf{k}'$ can be determined, we conclude that the instantaneous formulas (7)-(8) are only valid for $\omega \rightarrow \infty$, i.e. in the \emph{eikonal limit}.  Alternatively, the hypothesis of locality may be employed for the instantaneous measurement of the electromagnetic field
\begin{equation}
F_{(\alpha)(\beta)}=F_{\mu\nu}\Lambda^{\mu}_{\;\;(\alpha)}\Lambda^{\nu}_{\;\;(\beta)},
\end{equation}
which could then be subjected to the nonlocal process of Fourier analysis to reveal the frequency and wave vector content of the field.  This procedure will be illustrated using concrete examples in Section 2.  In Section 3 we consider the field equations for $F'_{\mu\nu}$, which has the interpretation of the Faraday tensor in the curvilinear coordinate system.  According to the fundamental observers in the rotating frame, the observables are still 
\begin{equation}
F_{(\alpha)(\beta)}=F'_{\mu\nu}\Lambda'^{\mu}_{\;\;(\alpha)}\Lambda'^{\nu}_{\;\;(\beta)},
\end{equation}
so that $F'_{\mu\nu}$ is an auxiliary field.  That is, $F_{\mu\nu}$ in (9) is the field measured by the fundamental static inertial observers, but $F'_{\mu\nu}$ does not have a similar significance.  The phenomena associated with spin-rotation-gravity coupling are briefly described in Section 4, which also contains a discussion of some of the observational consequences of spin-gravity coupling.  Finally, we summarize our results and conclusions in Section 5.
\section{Measurements of rotating observers}
Let us consider the frequency and wave vector content of electromagnetic radiation on the basis of Fourier analysis of the measured field as described by the local equation (9).  We assume that the plane monochromatic radiation field is given in the intertial frame by
\begin{equation}
{\bf{E}}(t,{\bf{r}})=A({\bf \hat {y}}\pm i{\bf \hat {n}})e^{-i\omega (t-{\bf \hat {k}} \cdot {\bf {r}})} 
\end{equation}
and ${\bf{B}}={\bf \hat{k}}\times{\bf{E}}$, where $A$ is a constant complex amplitude and 
\begin{equation}
{\bf \hat{k}}=\sin{\theta}\; {\bf \hat{x}}+\cos{\theta}\; {\bf \hat{z}}, \; \space {\bf \hat{n}}={\bf \hat{k}}\times{\bf \hat{y}}=-\cos{\theta}\;{\bf \hat{x}}+\sin{\theta}\;{\bf \hat{z}}.
\end{equation}
The upper (lower) sign in equation (11) indicates a positive (negative) helicity wave.  We concentrate on the fiducial observer $D$ at ${\bf{r}}=0$ and compute $F_{(\alpha)(\beta)}$.  It turns out that in this case ${\bf{E}}$ and ${\bf{B}}$ simply transform as three-vectors under a rotation.  It is therefore sufficient to consider the electric field measured by $D$, i.e. $E_{(i)}=-F_{(0)(i)}$,
\begin{eqnarray}
E_{(1)} = {\bf{E}}\cdot{\bf \hat {x'}} &=& E_{1}\cos{\Omega t}+E_{2}\sin{\Omega t}, \nonumber \\
E_{(2)} = {\bf{E}}\cdot{\bf \hat {y'}} &=& -E_{1}\sin{\Omega t}+E_{2}\cos{\Omega t}, \\
E_{(3)} = {\bf{E}}\cdot{\bf \hat {z'}} &=& E_{3}. \nonumber
\end{eqnarray}
Thus we find that
\begin{eqnarray}
E_{(1)} &=& A(\sin{\Omega t}\mp i\cos{\theta}\cos{\Omega t})e^{-i\omega t}, \nonumber \\
E_{(2)} &=& A(\cos{\Omega t}\pm i\cos{\theta}\sin{\Omega t})e^{-i\omega t}, \\
E_{(3)} &=& \pm iA \sin{\theta}\;e^{-i\omega t}. \nonumber
\end{eqnarray}
The Fourier analysis of these equations over all time results in $\omega '=\omega-m\Omega$, where $m=0,\pm 1$.  That is, $E_{(1)}$ and $E_{(2)}$ contain the frequency $\omega '=\omega -\Omega$ with amplitudes $(m=+1)$
\begin{equation}
\Psi'_{+(1)}=-\frac{1}{2}iA(1\pm \cos{\theta}),\;\;\Psi'_{+(2)}=\frac{1}{2}A(1\pm \cos{\theta}),
\end{equation}
respectively.  For $m=-1$, $E_{(1)}$ and $E_{(2)}$ contain the frequency $\omega '=\omega+\Omega$ with amplitudes
\begin{equation}
\Psi'_{-(1)}=\frac{1}{2}iA(1\mp \cos{\theta}),\;\;\Psi'_{-(2)}=\frac{1}{2}A(1\mp \cos{\theta}),
\end{equation}
respectively.  For $m=0$, $E_{(3)}$ contains the frequency $\omega'=\omega$ with amplitude $\pm iA\sin{\theta}$.

Comparing $\omega'=\omega-m\Omega$ with equation (7), which implies that \newline $\omega'($Doppler$)=\omega$, we recognize that our result approaches the Doppler formula as $\Omega/\omega \rightarrow 0$, i.e. in the ray limit.  This circumstance is entirely analogous to the quasi-classical approximation in wave mechanics.  In this connection, let us compute the average frequency measured by the observer $D$,
\begin{equation}
<\omega'>\;\;=\frac{\Sigma\;\omega'|\Psi'|^{2}}{\Sigma|\Psi'|^{2}}=(\omega-\Omega)P_{+}+\omega P_{0}+(\omega+\Omega)P_{-},
\end{equation}
where the summations contain five terms each involving the amplitudes mentioned above.  We find that
\begin{eqnarray}
\Sigma|\Psi'|^{2} &=& 2|A|^{2}, \nonumber \\
\Sigma\;\omega'|\Psi'|^{2} &=& 2|A|^{2}(\omega \mp \Omega\cos{\theta}), \nonumber \\
P_{+} &=& \frac{1}{4}(1\pm\cos{\theta})^{2}, \\
P_{0} &=& \frac{1}{2}\sin^{2}{\theta}, \nonumber \\
P_{-} &=& \frac{1}{4}(1\mp\cos{\theta})^{2},\nonumber
\end{eqnarray}
so that $<\omega'>\;=\omega \mp\Omega\cos{\theta}$.

It is important to provide proper physical interpretations for $P_{0}$ and $P_{\pm}$.  To this end, let us recall that for a particle of spin $\hbar$, the eigenstates of the particle with respect to the coordinate system $({\bf{\hat{y}}},{\bf{\hat{n}}},{\bf{\hat{k}}})$ can be transformed to the $({\bf{\hat{x}}},{\bf{\hat{y}}},{\bf{\hat{z}}})$ system using the matrix $(\mathcal{D}^{j}_{mm'})$ for $j=1$ given by [11]
\newline
\begin{equation}
\left[ \begin{array}{ccc} \frac{1}{2}(1+\cos{\theta}) & -\frac{1}{\sqrt{2}}\sin{\theta} & \frac{1}{2}(1-\cos{\theta}) \\ \frac{1}{\sqrt{2}}\sin{\theta} & \cos{\theta} & -\frac{1}{\sqrt{2}}\sin{\theta} \\\frac{1}{2}(1-\cos{\theta}) & \frac{1}{\sqrt{2}}\sin{\theta} & \frac{1}{2}(1+\cos{\theta}) \end{array} \right] .
\end{equation}
\newline
Thus a photon with definite helicity has an amplitude proportional to
\begin{equation}
\left[ \begin{array}{c} \frac{1}{2}(1\pm\cos{\theta})\\ \pm\frac{1}{\sqrt{2}}\sin{\theta}\\ \frac{1}{2}(1\mp\cos{\theta}) \end{array} \right],
\end{equation}
where the upper (lower) sign refers to positive (negative) helicity.  This result is simply obtained from the application of (19) on the helicity states and therefore corresponds to the first and third columns of (19) for positive and negative helicity states, respectively.  Thus the probability that a photon of definite helicity has spin $\hbar$ along the direction of rotation of the observer, i.e. $m=1$, is $P_{+}$ from (20) and (18).  Similarly for spin zero along the $z$-axis, i.e. $m=0$, we get $P_{0}$ and for spin $-\hbar$ along the $z$-axis, i.e. $m=-1$, we get from (20) the same result as $P_{-}$ given in (18).  It follows from a comparison of these results with (17) that in the formula $\omega'=\omega-m\Omega$, $m\hbar$ represents the projection of photon spin along the rotation axis of the observer, so that there is a coupling between the spin of the photon and the rotation of the observer [12].

Using the helicity vector ${\bf\hat{H}}=\pm{\bf\hat{k}}$, the average frequency measured by the observer $D$ is $<\omega'>\;=\omega-{\bf\hat{H}}\cdot{\bf{\Omega}}$, which again illustrates the phenomenon of helicity-rotation coupling.  Based on the quasi-classical derivation given here, this result is expected to hold in the eikonal approximation $\Omega/\omega<<1$; moreover, it can be shown [13,14] that this result follows directly from (14) if we assume that the time scale for observations in the noninertial frame --- which must necessarily extend over many periods of the wave --- is much shorter than $\Omega^{-1}$.

Let us next consider the wave vector measured by the fiducial observer $D$.  For this purpose, we need to consider an extended region, i.e. a cylindrical neighborhood around $D$ of radius $\rho_{0}<c/\Omega$, and compute the Fourier integral
\begin{equation}
{\hat{F}}_{(\alpha)(\beta)}(t,{\bf{k'}})=\int F_{(\alpha)(\beta)}(t,{\bf{r'}})e^{-i{\bf{k}}'\cdot{\bf{r}}'}d^{3}r'
\end{equation}
at a given time $t$.  The integration is carried out over the region with $0\leq\rho<\rho_{0}$, $0\leq\varphi<2\pi$ and $-\infty<z<\infty$.  We note here the operational problem of synchronization among the fundamental observers; this is related to the Sagnac effect [15] that is proportional to the cross-sectional area of the cylindrical region.  Moreover, it is important to recognize that with data given over length $L$, only wavelengths less than $L$ can be determined with any accuracy [16].  Since the fundamental observers employed in (21) are confined within the cylinder of radius $\rho_{0}$, our calculations would make sense for $|k'_{x}|>\rho^{-1}_{0}$ and $|k'_{y}|>\rho^{-1}_{0}$.

Computing the Faraday tensor $F_{(\alpha)(\beta)}$ as measured by the fundamental observers, we find using (9) that in general
\begin{eqnarray}
E_{(1)} &=& \gamma (E_{1}\cos{\phi}+E_{2}\sin{\phi})+\beta\gamma B_{3}, \nonumber \\
E_{(2)} &=& -E_{1}\sin{\phi}+E_{2}\cos{\phi}, \nonumber \\
E_{(3)} &=& \gamma E_{3}-\beta\gamma(B_{1}\cos{\phi}+B_{2}\sin{\phi}), \nonumber \\
B_{(1)} &=& \gamma (B_{1}\cos{\phi}+B_{2}\sin{\phi})-\beta\gamma E_{2}, \\
B_{(2)} &=& -B_{1}\sin{\phi}+B_{2}\cos{\phi}, \nonumber \\
B_{(3)} &=& \gamma B_{3}+\beta\gamma (E_{1}\cos{\phi}+E_{2}\sin{\phi}). \nonumber 
\end{eqnarray}
In the inertial frame ${\bf{E}}$ is given by (11) and ${\bf{B}}=\mp i\;{\bf{E}}$.  Evaluating (22), we find that it is sufficient for the rest of this discussion to concentrate on the electric field
\begin{equation}
E_{(i)}=Ae_{(i)}e^{-i\omega t+i{\bf{k}}\cdot{\bf{r}}},
\end{equation}
where $e_{(i)}$ are given by
\begin{eqnarray}
e_{(1)} &=& \gamma(\mp i\cos{\theta}\cos{\phi}+\sin{\phi}+\beta\sin{\theta}), \nonumber \\
e_{(2)} &=& \pm i\cos{\theta}\sin{\phi}+\cos{\phi}, \\
e_{(3)} &=& \gamma(\pm i\sin{\theta}+\beta\cos{\theta}\cos{\phi}\pm i\beta\sin{\phi}). \nonumber 
\end{eqnarray}
The Fourier integral (21) then reduces to
\begin{equation}
{\hat{E}}_{(i)}(t,{\bf{k'}})=2\pi A\;\delta(k_{z}-k'_{z})\;\;e^{-i\omega t}{\hat{e}}_{(i)}.
\end{equation}
Here ${\hat{e}}_{(i)}$ is given by
\begin{equation}
{\hat{e}}_{(i)}=\int_{0}^{\rho_{0}}\int_{0}^{2\pi} e_{(i)}\;e^{i\rho u \cos{(\varphi+\nu)}}\rho \; d\rho \; d\varphi,
\end{equation}
where $u=\sqrt{u_{x}^{2}+u_{y}^{2}}$,
\begin{eqnarray}
u_{x}=u\cos\nu &=& k_{x}\cos{\Omega t}+k_{y}\sin{\Omega t}-k'_{x}, \\
u_{y}=-u\sin\nu &=& -k_{x}\sin{\Omega t}+k_{y}\cos{\Omega t}-k'_{y}. 
\end{eqnarray}
The integration over $\varphi$ in (26) can be carried out using
\begin{eqnarray}
\int_{0}^{2\pi} e^{i\rho u \cos{(\varphi+\nu)}}d\varphi &=& 2\pi J_{0}(\rho u), \\
\int_{0}^{2\pi} \cos{\phi}\; e^{i\rho u \cos{(\varphi+\nu)}}d\varphi &=& 2\pi i\cos{(\Omega t -\nu)} J_{1}(\rho u), \\
\int_{0}^{2\pi} \sin{\phi}\; e^{i\rho u \cos{(\varphi+\nu)}}d\varphi &=& 2\pi i\sin{(\Omega t -\nu)} J_{1}(\rho u), 
\end{eqnarray}
where $J_{n}(\rho u)$ is the Bessel function of order $n$.  It is clear from (25) that $k'_{z}=k_{z}$ as measured by $D$.  Moreover, in close analogy with (17), we define the average of any function $Q(k'_{x},k'_{y})$ measured by $D$ as
\begin{equation}
<Q(k'_{x},k'_{y})>\;=\frac{\int Q(k'_{x},k'_{y})\mathcal{W}dk'_{x}dk'_{y}}{\int \mathcal{W} dk'_{x}dk'_{y}},
\end{equation}
where ${\mathcal{W}}=|{\hat{e}}_{(1)}|^2+|{\hat{e}}_{(2)}|^2+|{\hat{e}}_{(3)}|^2.$

To carry out the remaining integrations, we make a further simplifying assumption, namely, that $\rho_{0}\ll c/\Omega$; this is consistent with the eikonal approximation.  It then turns out that to lowest order, i.e. neglecting terms proportional to $\beta\ll 1$ in (24) and setting $\gamma\approx 1$, $\mathcal{W}$ is only a function of $u$.  Thus $<u_{x}>\;=\;<u_{y}>\;=0$, so that $<k'_{x}>\;={\bf{k}}\cdot{\bf{\hat{x'}}}$ and $<k'_{y}>\;={\bf{k}}\cdot{\bf{\hat{y'}}}$ using (27) and (28), respectively.  We therefore conclude that in this eikonal approximation $<{\bf{k'}}>\;={\bf{k}}$ and $<\omega'>\;=\omega-{\bf{\hat{H}}}\cdot{\bf{\Omega}}$ for the observer $D$.  Applying a somewhat extended form of the hypothesis of locality, we may argue that the measurement of any fundamental observer in the rotating frame is approximately related to that of $D$ by a Lorentz transformation [13].  It follows that for such a fundamental observer with velocity ${\bf{v}}$,
\begin{eqnarray}
<\omega'>\;=\gamma[(\omega-{\bf{\hat{H}}}\cdot{\bf{\Omega}})-{\bf{v}}\cdot{\bf{{k}}}], \\
<{\bf{k'}}>\;={\bf{k}}+\frac{1}{v^{2}}(\gamma -1)({\bf{v}}\cdot{\bf{k}}){\bf{v}}-\frac{1}{c^{2}}\gamma(\omega-{\bf{\hat{H}}}\cdot{\bf{\Omega}}){\bf{v}}.
\end{eqnarray}
The physical consequences of these results, i.e. the modified Doppler and aberration formulas, have been studied for interferometry with polarized light in rotating frames, Doppler tracking of spacecraft, etc. in [13,14].  Moreover, the observational evidence for helicity-rotation coupling in the radio, microwave and optical domains is presented in [17,14].

Our analysis in this section started with the Fourier decomposition of the field (14) measured by the observer $D$.  In fact, the analysis can be repeated for any uniformly rotating observer with the result that $\omega'=\gamma (\omega -M\Omega)$, where $M=0,\pm 1,\pm 2, ...$, is such that $\hbar M$ is the $z$-component of the \emph{total} angular momentum ${\bf{J}}$ of the radiation field.  In the eikonal approximation ${\bf{J}}={\bf{L}}+{\bf{S}}$, where ${\bf{L}}=\hbar {\bf{r}}\times{\bf{k}}$ and ${\bf{S}}=\hbar {\bf{\hat{H}}}$, so that with ${\bf{v}}={\bf{\Omega}}\times{\bf{r}}$ we recover equation (33) in the eikonal approximation.  Let us remark here that (33) and (34) are intermediate relations between the standard instantaneous Doppler and aberration formulas (7)-(8) and the results of Fourier analysis based on extended intervals of time and space.  Therefore, (33) is valid for observations that extend over an interval of time $\ll \Omega^{-1}$, while (34) is valid for measurements over a cylindrical region around the observer with radius $\ll c/\Omega$.

The general formula $\omega'=\gamma(\omega-M\Omega)$ has a consequence that must be pointed out: $\omega'=0$ if $\omega=M\Omega$, i.e. the electromagnetic wave can stand completely still for the whole set of fundamental observers.  That is, by a mere rotation the incident monochromatic radiation can be rendered static.  This circumstance is contrary to the spirit of relativity theory, where light travels with speed $c$ regardless of the motion of inertial observers. In fact, there is no observational evidence to date that any noninertial observer can be comoving with an electromagnetic wave.  In view of this situation, a nonlocal theory of accelerated observers has been developed that takes the past history of the observer into account and is based on the assumption that a basic radiation field can never stand completely still with respect to any observer [18-20].  To illustrate this nonlocal theory in the case of electrodynamics, we consider an observer that moves uniformly parallel to the $y$-axis with $x=r$ and speed $c\beta=r\Omega$ for $t<0$, but at $t=0$ starts uniform rotation with frequency $\Omega$ on a circle of radius $r$ in the $(x,y)$-plane.  The angle of rotation after time $t>0$ is given by $\phi=\Omega t=\gamma\Omega\tau$, where $\tau$ is the proper time of the accelerated observer.  According to the nonlocal theory, the field measured by the observer is given by (22) plus nonlocal contributions (that vanish as $\Omega/\omega \rightarrow 0$) as follows
\begin{eqnarray}
\mathcal{E}_{1}=E_{(1)}+\gamma^{2}\Omega\int^{\tau}_{0}(\sin{\phi'}E_{1}-\cos{\phi'}E_{2})\;d\tau', \nonumber \\
\mathcal{E}_{2}=E_{(2)}+\gamma\Omega\int^{\tau}_{0}(\cos{\phi'}E_{1}+\sin{\phi'}E_{2})\;d\tau', \nonumber \\ 
\mathcal{E}_{3}=E_{(3)}-\beta\gamma^{2}\Omega\int^{\tau}_{0}(\sin{\phi'}B_{1}-\cos{\phi'}B_{2})\;d\tau', \nonumber \\
\mathcal{B}_{1}=B_{(1)}+\gamma^{2}\Omega\int^{\tau}_{0}(\sin{\phi'}B_{1}-\cos{\phi'}B_{2})\;d\tau', \\
\mathcal{B}_{2}=B_{(2)}+\gamma\Omega\int^{\tau}_{0}(\cos{\phi'}B_{1}+\sin{\phi'}B_{2})\;d\tau', \nonumber \\
\mathcal{B}_{3}=B_{(3)}+\beta\gamma^{2}\Omega\int^{\tau}_{0}(\sin{\phi'}E_{1}-\cos{\phi'}E_{2})\;d\tau'. \nonumber 
\end{eqnarray}
The physical predictions of nonlocal electrodynamics have been described in detail before (see [18]-[20] and the references cited therein); the theory agrees with all available observational data at present.  In particular, the general result for spin-rotation coupling, $\omega'=\gamma(\omega-M\Omega)$, is valid except for the case in which $\omega=M\Omega$ and according to the hypothesis of locality the radiation field becomes static and comoving with the uniformly rotating observers.  Using (35), however, we find that ${\bf{\mathcal{E}}}$ and ${\bf{\mathcal{B}}}$ are not static for $\omega'=0$ and behave as in the case of resonance.  Moreover, the nonlocal theory predicts that helicity-rotation coupling in general affects the amplitude of the measured field as well.  For instance, an observer rotating with frequency $\Omega$ in the positive (negative) sense about the direction of propagation of a plane positive-helicity wave of frequency $\omega \gg\Omega$ would find that the amplitude of the field is larger (smaller) by a factor that is given approximately by $1+\Omega/\omega$ $(1-\Omega/\omega)$.  It is important to test experimentally such a prediction of the nonlocal theory of accelerated observers.

\section{Wave propagation}
To describe the propagation of electromagnetic waves in a rotating frame of reference, it is useful to employ the Skrotskii formalism [20]-[25].  Let $F'_{\mu\nu}$ be the Faraday tensor in terms of curvilinear coordinates such that $F_{\mu\nu}dx^{\mu}\wedge dx^{\nu}=F'_{\rho\sigma}dx'^{\rho}\wedge dx'^{\sigma}$.  We then require that the coordinates be quasi-Cartesian; in fact, this is the case in (6) since $g'_{\mu\nu}\rightarrow \eta_{\mu\nu}$ as $\Omega\rightarrow 0$.  Now consider the standard decompositions $F'_{\mu\nu}\rightarrow({\bf{E'}},{\bf{B'}})$ and $\sqrt{-g'}F'^{\mu\nu}\rightarrow(-{\bf{D'}},{\bf{H'}})$, so that Maxwell's equations in the curvilinear coordinates --- i.e. $F'_{\mu\nu,\rho}+F'_{\nu\rho,\mu}+F'_{\rho\mu,\nu}=0$ and $(\sqrt{-g'}F'^{\mu\nu})_{,\nu}=0$ --- take their flat spacetime form but in the presence of a ``medium" with special constitutive properties characteristic of a gyrotropic medium.  That is,
\begin{eqnarray}
D'_{i}=\epsilon_{ij}E'_{j}-({\bf{G}}\times{\bf{H'}})_{i}, \\
B'_{i}=\mu_{ij}H'_{j}+({\bf{G}}\times{\bf{E'}})_{i}, 
\end{eqnarray}
where the dielectric and permeability tensors are equal,
\begin{equation}
\epsilon_{ij}=\mu_{ij}=\frac{\sqrt{-g'}\hspace{1ex}g'^{ij}}{-g'_{00}}
\end{equation}
and the gyrotropic vector ${\bf{G}}$ --- representing the rotational motion of the medium --- is given by
\begin{equation}
G_{i}=\frac{g'_{0i}}{-g'_{00}}.
\end{equation}

Let us next define the Kramers vectors ${\bf{F}}^{\pm}$ and ${\bf{S}}^{\pm}$ in terms of the complex fields ${\bf{E}}',{\bf{B}}',{\bf{D}}'$ and ${\bf{H}}'$ as follows
\begin{equation}
{\bf{F}}^{\pm}={\bf{E}}'\pm i{\bf{H}}',\;\;{\bf{S}}^{\pm}={\bf{D}}'\pm i{\bf{B}}'.
\end{equation}
The field equations may then be expressed as
\begin{equation}
{\bf{\nabla}}'\cdot{\bf{S}}^{\pm}=0,\;\;{\bf{\nabla}}'\times{\bf{F}}^{\pm}=\pm i\frac{\partial{\bf{S}}^{\pm}}{\partial t}
\end{equation}
and the constitutive relations become
\begin{equation}
S_{k}^{\pm}=\epsilon_{kj}F^{\pm}_{j}\pm i({\bf{G}}\times{\bf{F}}^{\pm})_{k}.
\end{equation}

The ``medium" under consideration here is stationary; therefore, it is useful to assume a temporal dependence of the form exp$(-i\omega t)$ with $\omega >0$ for all field quantities.  Here $\omega$ is the frequency of the mode under consideration.  Equations (41) and (42) then reduce to a single wave equation for ${\bf{F}}^{\pm}$,
\begin{equation}
{\bf{\nabla}}'\times{\bf{F}}^{\pm}=\pm \omega({\bf{\epsilon}}{\bf{F}}^{\pm}\pm i{\bf{G}}\times{\bf{F}}^{\pm}).
\end{equation}
It is important to recognize that in the absence of the ``medium", ${\bf{F}}^{+}$ is the complex amplitude for a wave of positive helicity; that is, it represents a right circularly polarized (RCP) wave.  On the other hand, ${\bf{F}}^{-}$ is the complex amplitude for a wave of negative helicity; that is, it represents a left circularly polarized (LCP) wave.  In what follows we will keep these designations even in the presence of the medium since they are recovered in the limit $\Omega\rightarrow 0$.

According to the hypothesis of locality, the actual field strength measured by the observer is given by the projection of the Faraday tensor on the orthonormal tetrad frame of the observer; therefore, in agreement with (10),
\begin{equation}
F_{(\alpha)(\beta)}'=F_{\mu\nu}'\Lambda'^{\mu}_{\;\;(\alpha)}\Lambda'^{\nu}_{\;\;(\beta)}=F_{(\alpha)(\beta)},
\end{equation}
where $\Lambda'^{\mu}_{\;\;(\alpha)}$ is given by (5).  We find that
\begin{eqnarray}
E_{(1)} &=& \gamma (E'_{1}\cos{\varphi}+E'_{2}\sin{\varphi}), \nonumber \\
E_{(2)} &=& -E'_{1}\sin{\varphi}+E'_{2}\cos{\varphi}, \nonumber \\
E_{(3)} &=& \gamma E'_{3}, \nonumber \\
B_{(1)} &=& \gamma^{-1} (B'_{1}\cos{\varphi}+B'_{2}\sin{\varphi})-\beta\gamma E'_{3}, \\
B_{(2)} &=& -B'_{1}\sin{\varphi}+B'_{2}\cos{\varphi}, \nonumber \\
B_{(3)} &=& \gamma^{-1} B'_{3}+\beta\gamma (E'_{1}\cos{\varphi}+E'_{2}\sin{\varphi}). \nonumber 
\end{eqnarray}
The nonlocal aspects of the measurement process may be taken into account as in (35).

It follows from the form of the metric (6) that $\sqrt{-g'}=1$ and $g'^{ij}=\delta_{ij}-({\bf{\Omega}} \times {\bf{x}}')_{i}({\bf{\Omega}} \times {\bf{x}}')_{j}$; therefore, (38) and (39) imply that in this case
\begin{equation}
{\bf{\epsilon}}=\gamma^{2}\left[ \begin{array}{ccc} 1-\Omega^{2}y'^{2} & \Omega^{2}x'y' & 0 \\ \Omega^{2}x'y' & 1-\Omega^{2}x'^{2} & 0 \\0 & 0 & 1 \end{array} \right] 
\end{equation}
and
\begin{equation}
{\bf{G}}=\gamma^{2}{\bf{\Omega}}\times{\bf{x}}',
\end{equation}
where $\gamma^{-2}=1-\Omega^{2}(x'^{2}+y'^{2})$.  Introducing cylindrical coordinates $(\rho,\varphi,z)$ such that $x'=\rho\cos{\varphi}, y'=\rho\sin{\varphi}$ and $z'=z$, we find that (43) can be written as
\begin{eqnarray}
\frac{1}{\rho}\frac{\partial F^{\pm}_{z}}{\partial\varphi}-\frac{\partial F^{\pm}_{\varphi}}{\partial{z}}- i\omega\beta\gamma^{2}F^{\pm}_{z}=\pm\omega\gamma^{2}F^{\pm}_{\rho}, \nonumber \\
\frac{\partial F^{\pm}_{\rho}}{\partial z}-\frac{\partial F^{\pm}_{z}}{\partial \rho}=\pm\omega F^{\pm}_{\varphi}, \\
\frac{1}{\rho}\frac{\partial}{\partial\rho}(\rho F^{\pm}_{\varphi})-\frac{1}{\rho}\frac{\partial F^{\pm}_{\rho}}{\partial\varphi}+ i\omega\beta\gamma^{2}F^{\pm}_{\rho}=\pm\omega\gamma^{2}F^{\pm}_{z}.\nonumber
\end{eqnarray}

Let us look for a solution of these equations that is of the form
\begin{equation}
{\bf{F^{\pm}}}={\mathcal{A}}_{\pm}\;e^{-i\omega t+ikz}e^{im\varphi}\;{\bf{f}}^{\pm},
\end{equation}
where $\mathcal{A}_{+}$ and $\mathcal{A}_{-}$ are constant amplitudes, ${\bf{f^{\pm}}}$ depends only on the radial coordinate $\rho$, $m$ is an integer and $k$ is a constant wave number.  The system (48) then reduces to
\begin{eqnarray}
\pm a\gamma^{2}f^{\pm}_{\rho}+ibf^{\pm}_{\varphi}-i(\frac{m}{\beta}-a\beta\gamma^{2})f^{\pm}_{z}=0, \nonumber \\
ibf^{\pm}_{\rho}\mp af^{\pm}_{\varphi}-\frac{d}{d\beta}f^{\pm}_{z}=0,  \\
i(\frac{m}{\beta}-a\beta\gamma^{2})f^{\pm}_{\rho}-\frac{1}{\beta}\frac{d}{d\beta}(\beta f^{\pm}_{\varphi})\pm a\gamma^{2}f^{\pm}_{z}=0, \nonumber
\end{eqnarray}
where $a=\omega/\Omega$ and $b=k/\Omega$.  In the first two equations of (50), $f^{\pm}_{\rho}$ and $f^{\pm}_{\varphi}$ can be expressed in terms of $f^{\pm}_{z}$ and $\frac{df^{\pm}_{z}}{d\beta}$; then, the substitution of these relations in the last equation results in a single second-order linear differential equation for $f^{\pm}_{z}$,
\begin{equation}
\frac{d^{2}f^{\pm}_{z}}{d\beta^{2}}+\frac{1}{\beta}\;\frac{a^{2}-b^{2}-b^{2}\beta^{2}}{a^{2}-b^{2}+b^{2}\beta^{2}}\;\frac{df^{\pm}_{z}}{d\beta}+\mathcal{V}_{\pm}(\beta)f^{\pm}_{z}=0,
\end{equation}
where $\mathcal{V}_{\pm}$ is given by
\begin{equation}
\mathcal{V}_{\pm}=a^{2}\gamma^{2}-b^{2}-\frac{1}{\beta^{2}\gamma^{2}}(m-a\beta^{2}\gamma^{2})^{2}\mp 2b\gamma^{2}\;\frac{a^{2}+ma-b^{2}}{a^{2}\gamma^{2}-b^{2}}.
\end{equation}
In the physical region $0\leq\beta<1$, (51) has regular singularities at $\beta=0$, $\sqrt{1-a^{2}/b^{2}}$ for $a^{2}<b^{2}$ and $\beta=1$; therefore, one can in general find series solutions around any point using Frobenius's method provided that such series solutions are convergent over the region of interest.

Let us note that near the axis of rotation, $\beta\rightarrow0$, the behavior of $f^{\pm}_{z}$ can be determined from (51)-(52) to be $f^{\pm}_{z}\sim\beta^{\;p}$, where $p^{2}=m^{2}$.  This is exactly the same as in the familiar case of the solution of the Schr\"{o}dinger equation for an electron in a constant magnetic field.  We choose $p=|m|$, so that $f^{\pm}_{z}$ is regular on the axis of rotation.

It is important to put equation (51) in a standard form.  To this end, let us first assume that $b=0$.  Then (51) takes the form of Bessel's equation
\begin{equation}
\beta^{2}\frac{d^{2}f^{\pm}_{z}}{d\beta^{2}}+\beta\frac{df^{\pm}_{z}}{d\beta}+[(a+m)^{2}\beta^{2}-m^{2}]f^{\pm}_{z}=0.
\end{equation}
Thus for radiation propagating orthogonal to the axis of rotation, the solutions are of the form $f^{\pm}_{z}=J_{m}(\hat{\beta})$, where $\hat{\beta}=(a+m)\beta$.

Let us next assume that $a^{2}=b^{2}\neq 0$, corresponding to a wave propagating along the axis of rotation with $\omega^{2}=k^{2}$.  Then, equation (51) takes the form
\begin{equation}
\beta^{2}\frac{d^{2}f^{\pm}_{z}}{d\beta^{2}}-\beta\frac{df^{\pm}_{z}}{d\beta}+m[(m+2a)\beta^{2}-(m\pm 2\frac{b}{a})]f^{\pm}_{z}=0.
\end{equation}
For $m=0$, (54) has the simple solution $f^{\pm}_{z}=\frac{1}{2}\beta^{2}+K_{\pm}$, where $K_{+}$ and $K_{-}$ are integration constants.  Equations (50) can then be used to calculate $f^{\pm}_{\rho}$ and $f^{\pm}_{\varphi}$.  To ensure that these functions do not diverge on the axis $(\beta=0)$, we must choose $K_{\pm}=\pm \Omega/k$.  It then follows that
\begin{equation}
f^{\pm}_{\rho}=-i(\pm \frac{1}{2}+\frac{\Omega}{k})\beta,\; f^{\pm}_{\varphi}=\frac{1}{2}\frac{k}{\omega}\beta,\; f^{\pm}_{z}=\frac{1}{2}\beta^{2}\pm \frac{\Omega}{k}
\end{equation}
is a solution of (50) that disappears for $\Omega\rightarrow 0$.  We therefore have a solution of Maxwell's equation in the rotating frame once we choose $\mathcal{A}_{+}$ and $\mathcal{A}_{-}$ in (49); for the sake of simplicity, we consider a positive helicity wave with $\mathcal{A}_{+}=1$ and $\mathcal{A}_{-}=0$.  It is interesting to determine in this case the corresponding fields ${\bf{E}}$ and ${\bf{B}}$ measured by the static inertial observers in the background global frame.  A detailed calculation shows that in the inertial frame the electromagnetic field has positive helicity as well and is given by the real part of 
\begin{equation}
{\bf{E}} = \frac{\Omega}{4k}\left[ \begin{array}{c} -ikx-\omega y \\ \omega x-iky \\2 \end{array} \right]e^{-i\omega t+ikz}
\end{equation}
and ${\bf{B}}=-i{\bf{E}}$.  We note that in this case these free fields in the inertial frame with $\omega^{2}=k^{2}$ are divergent for $\rho \rightarrow \infty$.

Finally, we consider the general case with $b\neq0$ and $a^{2}\neq b^{2}$.  Let $f^{\pm}_{z}=\beta^{|m|}\psi_{\pm}(\xi)$, where $\xi=(1-a^{2}/b^{2})^{-1}\beta^{2}$.  Then, (51) implies that
\begin{equation}
\xi(1-\xi)\psi''_{\pm}+(1+|m|)(1-\xi)\psi'_{\pm}+[C(1-\xi)\pm N]\psi_{\pm}=0,
\end{equation}
where a prime denotes differentiation with respect to $\xi$ and
\begin{equation}
C=\frac{1}{4}(a^{2}-b^{2})[1-(\frac{a+m}{b})^{2}],\; N=\frac{1}{2b}(a^{2}+ma-b^{2}).
\end{equation}

It follows from the form of equation (57) that $f^{+}_{z}$ and $f^{-}_{z}$ propagate differently in a rotating frame of reference.  The helicity-rotation coupling is thus hidden in the solutions of these equations, which, to our knowledge, cannot in general be simply expressed in terms of the special functions of mathematical physics.

\section{Spin-rotation-gravity coupling}
It turns out that the phenomenon of helicity-rotation coupling is an instance of a general effect [26]-[27].  For the investigation of phenomena in a laboratory fixed on the rotating Earth, the Hamiltonian must be augmented by the spin-rotation-gravity term
\begin{equation}
H\approx-{\bf{S}}\cdot{\bf{\Omega}}+{\bf{S}}\cdot{\bf{\Omega}}_{P},
\end{equation}
where ${\bf{\Omega}}_{P}$ is the gravitomagnetic precession frequency of an ideal gyroscope with its center of mass at rest in the laboratory and is given approximately by
\begin{equation}
{\bf{\Omega}}_{P}=\frac{GJ}{c^{2}r^{3}}[3({\bf{\hat{r}}}\cdot{\bf{\hat{J}}}){\bf{\hat{r}}}-{\bf{\hat{J}}}].
\end{equation}
It follows from (59) and the dependence of ${\bf{\Omega}}_{P}$ on position that there must be an analogue of the Stern-Gerlach force
\begin{equation}
{\bf{\mathcal{F}}}=-{\bf{\nabla}}({\bf{S}}\cdot{\bf{\Omega}}_{P}),
\end{equation}
which is of gravitational origin and depends only on the spin of the particle.  In the correspondence limit, this force can be deduced from the Mathisson-Papapetrou spin-curvature force [26].  We note that classical spin is generally proportional to mass, whereas in the quantum theory spin is simply proportional to $\hbar$.
It follows from (61) that a neutron with spin up falls differently in the gravitational field of the Earth than a neutron with spin down.  This equivalence-principle violating effect is very small [17, 26]; due to its fundamental nature, however, it is interesting to illustrate this situation by computing the differential deflection angle of particles with spin in the exterior gravitational field of a rotating source as in Figure 2.
\begin{figure}
\includegraphics{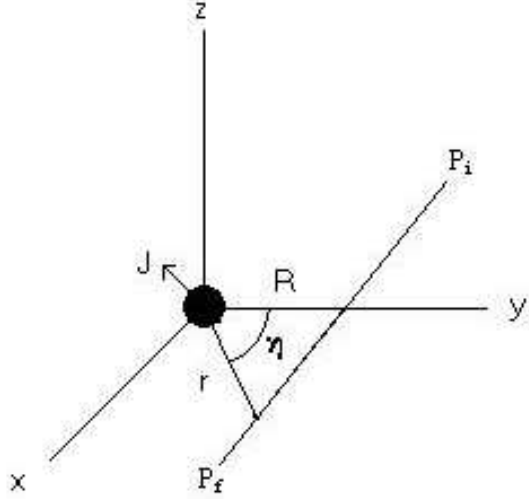}
\caption{A free particle travels from an initial point $P_{i}$ to a final point $P_{f}$ parallel to the $x$-axis in the ($x$,$y$)-plane.  This path will be deflected in the presence of a rotating gravitational source located at the origin}
\end{figure}
We assume that the particle starts from point $P_{i}$ and travels to point $P_{f}$ on a straight line with constant speed $v$; that is, to illustrate the $\emph{differential}$ deflection we neglect the usual deflection due to the attraction of gravity.  In fact, the differential deflection evaluated below is expected to occur around the deflected path.

We are interested in computing the impulse
\begin{equation}
{\bf{I}}=\int^{P_{f}}_{P_{i}}{\bf{\mathcal{F}}}\;dt,
\end{equation}
\newline
\newline
where
\begin{equation}
{\bf{\mathcal{F}}}=\frac{3GJ}{c^{2}r^{4}}[5({\bf{S}}\cdot{\bf{\hat{r}}})({\bf{\hat{J}}}\cdot{\bf{\hat{r}}}){\bf{\hat{r}}}-({\bf{S}}\cdot{\bf{\hat{J}}}){\bf{\hat{r}}}-({\bf{S}}\cdot{\bf{\hat{r}}}){\bf{\hat{J}}}- ({\bf{\hat{J}}}\cdot{\bf{\hat{r}}}){\bf{S}}].
\end{equation}
Without any loss in generality, we may assume that the trajectory of the particle (or ray) is parallel to the $x$-axis with impact parameter $R$ in the ($x$,$y$)-plane as in Figure 2.  A detailed calculation reveals that the transverse impulse due to the spin-gravity force is given by
\begin{eqnarray}
I_{y} &=&\frac{GJ}{c^{2}vR^{3}}[-(3s_{1}-6s_{3}+3s_{5})S_{x}\hat{J}_{x}+(6s_{1}-7s_{3}+3s_{5})S_{y}\hat{J}_{y} \nonumber \\ 
& & \mbox{}-(3s_{1}-s_{3})S_{z}\hat{J}_{z}+(c_{3}-3c_{5})(S_{x}\hat{J}_{y}+S_{y}\hat{J}_{x})],  
\end{eqnarray}
where
\begin{equation}
s_{n}=\sin^{n}{\eta_{f}}-\sin^{n}{\eta_{i}},\;\;c_{n}=\cos^{n}{\eta_{f}}-\cos^{n}{\eta_{i}}
\end{equation}
and the angle $\eta$ is defined as in Figure 2.  Similarly,
\begin{equation}
I_{z}=\frac{GJ}{c^{2}vR^{3}}[c_{3}(S_{x}{\hat{J}}_{z}+ S_{z}{\hat{J}}_{x})-(3s_{1}-s_{3})( S_{y}{\hat{J}}_{z}+ S_{z}{\hat{J}}_{y})].
\end{equation}
Moreover, the tangential impulse is given by
\begin{eqnarray}
I_{x}&=&\frac{-GJ}{c^{2}vR^{3}}[(2c_{3}-3c_{5})S_{x}{\hat{J}}_{x}+(-c_{3}+3c_{5}) S_{y}{\hat{J}}_{y}-c_{3}S_{z}{\hat{J}}_{z} \nonumber \\
& & \mbox{}+(3s_{1}-6s_{3}+3s_{5})( S_{x}{\hat{J}}_{y}+ S_{y}{\hat{J}}_{x})].
\end{eqnarray}

If the spin of the particle is in the direction of its motion, i.e. ${\bf{S}}=(S_{x},0,0)$, then integration from $\eta_{i}=-\pi/2$ to $\eta_{f}=\pi/2$ reveals that in the scattering case ${\bf{I}}=0$ and there is no differential deflection.  This is in agreement with the result of [28] for the case of electromagnetic radiation.  Next, let us consider the case of electromagnetic waves of definite helicity starting at the pole and propagating to infinity.  With ${\bf{\hat{J}}}=(0,1,0)$ and ${\bf{S}}=(\pm\hbar,0,0)$, we find $I_{x}=I_{z}=0$ and $I_{y}=\pm 2GJ\hbar/(c^{3}R^{3})$.  Thus positive (negative) helicity radiation is differentially deflected away from (toward) the source such that the angle of separation of the RCP and LCP waves within the wave packet is given by
\begin{equation}
\Delta=\frac{2GJ\lambda}{\pi c^{3}R^{3}}
\end{equation}
in agreement with [28].  If the radiation starts from (0,$R$,0) as in the previous case but ${\bf{\hat{J}}}=(0,\cos{\zeta},\sin{\zeta})$, then $I_{x}=0$, but
\begin{equation}
I_{y}=\pm\frac{2GJ\hbar}{c^{2}R^{3}}\cos{\zeta},\;\; I_{z}=\mp\frac{GJ\hbar}{c^{2}R^{3}}\sin{\zeta},
\end{equation}
so that we recover (68) for $\zeta=0$.

Finally, let us consider two beams propagating from $-\infty$ to $\infty$ on the same path but polarized parallel and antiparallel to ${\bf{J}}$ such that ${\bf{\hat{J}}}=(0,0,1)$ and ${\bf{S}}=\pm\frac{1}{2}\hbar{\bf{\hat{J}}}$.  Then, we find from (64)-(67) that $I_{x}=I_{z}=0$ and 
\begin{equation}
I_{y}=\mp \frac{2GJ\hbar}{c^{2}vR^{3}},
\end{equation}
so that the force is attractive (repulsive) when ${\bf{J}}$ and ${\bf{S}}$ are parallel (antiparallel) and the total deflection angle between the beams is $4GJ\hbar/(c^{2}vR^{3}P)$, where $P$ is the momentum of the particle.  As already pointed out in [25, 28], such effects are very small in the field of the Earth, but could possibly become significant in the field of a neutron star or a black hole.
\section{Discussion}
It has been the purpose of this paper to clarify various aspects of the phenomenon of helicity-rotation coupling.  This is of particular interest at present in connection with its applications in the Global Positioning System, where it is known as phase wrap-up [29], and the Doppler tracking of spacecraft [30].  In the process of communicating with artificial satellites, circularly polarized radiation is generally employed and the source and the receiver both rotate.  Therefore, the modifications in the standard Doppler and aberration formulas due to the coupling of photon spin with rotation should be taken into account in the analysis of satellite data. 
\newpage
\noindent {References}
\vspace{.15in}
\begin{description}
\item{[1]} F.W. Hehl and Y.N. Obukhov, Foundations of Classical Electrodynamics (Birkh\"{a}user, Boston, 2003)
\item{[2]} R. Neutze and G.E. Stedman, Phys. Rev. A 58(1998) 82
\item{[3]} Y.Q. Cai and G. Papini, Phys. Rev. Lett. 66 (1991) 1259
\item{[4]} J. Van Bladel, Relativity and Engineering (Springer, Berlin, 1984)
\item{[5]} E.J. Post, Rev. Mod. Phys. 39 (1967) 475
\item{[6]} C.V. Heer, Phys. Rev. 134 (1964) A799
\item{[7]} B. Mashhoon and U. Muench, Ann. Phys. (Leipzig) 11 (2002) 532
\item{[8]} L.H. Ryder and B. Mashhoon, in: Proc. Ninth Marcel Grossmann Meeting (Rome, 2000), edited by V.G. Gurzadyan, R.T. Jantzen and R. Ruffini (World Scientific, Singapore, 2002), pp. 486-497
\item{[9]} B. Mashhoon, Phys. Lett. A 143 (1990) 176
\item{[10]} B. Mashhoon, Phys. Lett. A 145 (1990) 147
\item{[11]} A.R. Edmonds, Angular Momentum in Quantum Mechanics (Princeton University Press, Princeton, 1960), p. 57
\item{[12]} B. Mashhoon, Found. Phys. (Wheeler Festschrift) 16 (1986) 619
\item{[13]} B. Mashhoon, Phys. Lett. A 139 (1989) 103
\item{[14]} B. Mashhoon, Phys. Lett. A 306 (2002) 66
\item{[15]} G.E. Stedman, Rep. Prog. Phys. 60 (1997) 615
\item{[16]} B. Mashhoon, Phys. Lett. A 122 (1987) 299
\item{[17]} B. Mashhoon, R. Neutze, M. Hannam and G.E. Stedman, Phys. Lett. A 249 (1998) 161
\item{[18]} B. Mashhoon, Phys. Rev. A 47 (1993) 4498
\item{[19]} C. Chicone and B. Mashhoon, Ann. Phys. (Leipzig) 11 (2002) 309
\item{[20]} C. Chicone and B. Mashhoon, Phys. Lett. A 298 (2002) 229
\item{[21]} G.V. Skrotskii, Dokl. Akad. Nauk USSR 114 (1957) 73 [Sov. Phys. - Dokl. 2 (1957) 226]
\item{[22]} J. Plebanski, Phys. Rev. 118 (1960) 1396
\item{[23]} F. de Felice, Gen. Rel. Grav. 2 (1971) 347
\item{[24]} A.M. Volkov, A.A. Izmest'ev and G.V. Skrotskii, Zh. Eksp. Teor. Fiz. 59 (1970) 1254 [Sov. Phys. - JETP 32 (1971) 686]
\item{[25]} B. Mashhoon, Phys. Rev. D 11 (1975) 2679
\item{[26]} B. Mashhoon, Class. Quantum Grav. 17 (2000) 2399
\item{[27]} G. Papini, in: Advances in the Interplay Between Quantum and Gravity Physics, edited by P.G. Bergmann and V. de Sabbata (Kluwer Academic Publishers, Dordrecht, 2002), pp. 317-338
\item{[28]} B. Mashhoon, Phys. Lett. A 173 (1993) 347
\item{[29]} N. Ashby, in: Proc. GR-15, Gravitation and Relativity: At the Turn of the Millennium, edited by N. Dadhich and J. Narlikar (Inter-University Center for Astronomy and Astrophysics, Pune, India, 1997), pp. 231-258
\item{[30]} J.D. Anderson et al., Phys. Rev. D 65 (2002) 082004
\end{description}
\end{document}